\documentstyle[prb,aps]{revtex}

\draft  

\input psfig.tex   

\begin{document}

\title{ Interaction of pulses in nonlinear Schr\"odinger model}
\author{E. N. Tsoy$^a$\footnote{Corresponding author:
  etsoy@physic.uzsci.net}  and F. Kh. Abdullaev$^{a,b}$}
\address{ $^a$ Physical-Technical Institute
  of the Uzbek Academy of Sciences,\\
  2-B, Mavlyanov street, Tashkent, 700084, Uzbekistan\\
  $^b$ Instituto de Fisica Teorica, UNESP, Sao Paulo, Brasil
}

\date{\today}
\maketitle

\begin{abstract}
  The interaction of two rectangular pulses in nonlinear
Schr\"odinger model is studied by solving the appropriate
Zakharov-Shabat system. It is shown that two real pulses may result
in appearance of moving solitons. Different limiting cases, such as
a single pulse with a phase jump, a single chirped pulse, in-phase
and out-of-phase pulses, and pulses with frequency separation, are
analyzed. The thresholds of creation of new solitons and
multi-soliton states are found.

(Submitted to Phys. Rev. E, 2003)
\end{abstract}


\section{Introduction}
\label{Sect:Intro}

   The nonlinear Schr\"odinger (NLS) equation is an important model
of the theory of modulational waves. It describes the propagation of
pulses in optical fibers \cite{Optics,Haseg}, the dynamics of laser
beams in a Kerr media, or the nonlinear difraction \cite{Zakh},
waves in plasma \cite{Ikezi}, and the evolution of a Bose-Einstein
condensate wave function \cite{GP}. The NLS equation is written in
dimensionless form as the following:
\begin{equation}
   i u_z + u_{xx}/2 + |u|^2 u = 0,\
\label{eq:NLS}
\end{equation}
where $u(x,z)$ is a slowly varying wave envelope, $z$  is the
evolutional variable, and $x$ is associated with the spatial
variable.

   An exact solution of the NLS equation has a form of a
soliton:
\begin{equation}
  u(x,z)= 2 \eta\; {\rm sech}[2 \eta(x+2\xi z-x_0)]
    \exp[-2i\xi x - 2i(\xi^2-\eta^2) z + i\phi_0 ]\ .
\label{eq:Solit}
\end{equation}
where $2\eta$ and $2\xi$ are amplitude, or the inverse width, and
the velocity of the soliton, $x_0$ and $\phi_0$ are the initial
position and phase, respectively. The soliton represents a basic
mode and plays a fundamental role in nonlinear processes. The
dynamics of NLS solitons and single pulses even in the presence of
various perturbations is well understood (see e.g. Refs.~
\cite{Optics,Haseg} and reference therein). However, the evolution
of several pulses is not studied in details. In recent works
\cite{Inter} (see also book~\cite{Haseg}) mostly an interaction of
solitons and near-soliton pulses was considered. A study of {\em
near-soliton} pulses, especially a use of the effective particle
approach, often results in small variation of soliton parameters,
including the soliton velocities and as a consequence weak repulsion
or attraction of solitons. Such a study do not involve a possibility
of appearance of additional solitons. However, for many applications
it is necessary to consider the interaction of pulses  with
arbitrary amplitudes or pulses with different parameters. For
example, in optical communication systems with the wavelength
division multiplexing (WDM), the initial signal consist of several
solitons with different frequencies. An estimation of the critical
separation between pulses is important for determination of the
repetition rate of a particular transmission scheme.

   In present work the interaction of two pulses in the NLS model is
studied both theoretically and numerically. We show the presence of
different scenarios of the behaviour, depending on the initial
parameters of the pulses, such as the pulse areas, the relative
phase shift, the spatial and frequency separations. One of our main
observation is a fact that a pure real initial condition of the NLS
equation can result in additional {\em moving} solitons. As a
consequence the number of solitons, emerging from two pulses
separated by some distance, can be larger than the sum of the
numbers of solitons, emerging from each pulse. Such properties were
also found for the Manakov system \cite{Abdul}, which is a vector
generalization of the NLS equation. The scalar NLS equation was
studied in \cite{Abdul} as a particular case. Later similar results
and approximation formulas for the soliton parameters were obtained
in papers \cite{Klau,Desaix} (see also \cite{Desai2}). In works
\cite{Abdul,Klau,Desaix,Desai2} mostly the interaction of {\em real}
pulses was analyzed, while here we consider pulses with non-zero
relative phase shift and frequency separation. A preliminary version
of this study was presented in work \cite{Tsoy}.

   The paper is organized as the following. The linear scattering
problem associated with the NLS equation is considered in
Section~\ref{Sect:Dir}. We also present the general solution of the
problem for the case of two rectangular pulses. In
Section~\ref{Sect:Res} we study different particular cases, such as
two in-phase pulses, two out-of-phase pulses, a single pulse with a
phase jump, a single chirped pulse, and two pulses with the
frequency separation. The results and conclusions are summarized in
Section \ref{Sect:Conc}.

\section{ Direct scattering problem }
\label{Sect:Dir}

   In this paper we are interested only in an asymptotic state of
the pulse interaction. In order to simplify the problem and to
obtain exact results we consider the interaction of two {\em
rectangular} pulses (``boxes''). Therefore we take the following
initial conditions for Eq.~(\ref{eq:NLS}):
\begin{equation}
  u(x,0) \equiv U(x) = \left\{ \begin{array} {ll}
     Q_1 \exp[2i \nu_1 x], & {\rm for}\ \   x_1 < x < x_2, \\
     Q_2 \exp[2i \nu_2 x], & {\rm for}\ \  x_3 < x < x_4, \\
     0 , & {\rm otherwise}\ ,
  \end{array} \right.
\label{eq:IC}
\end{equation}
where $Q_1$ and $Q_2$ are the  complex constant amplitudes, $w_1
\equiv x_2 - x_1$ and $w_2\equiv x_4-x_3$ are the pulse widths, and
$2\nu_1$ and $2\nu_2$ are the detunings.

   It is known that NLS equation is integrable by the inverse
scattering transform method \cite{Zakh}. As follows from this fact,
initial conditions, which decreases sufficiently fast at
$x= \pm \infty$, results in a set of solitons and linear waves (so
called, radiation). The number $N$ and parameters of solitons
emerging from an initial condition are found from the
solution of the Zakharov-Shabat scattering problem \cite{Zakh}:
\begin{eqnarray}
  i {\partial \psi_{1} \over \partial x } - i U(x)\, \psi_2 &=&
    \lambda \psi_1\ ,
\nonumber\\
  -i  {\partial \psi_{2} \over \partial x } - i U^{*}(x)\, \psi_1 &=&
    \lambda \psi_2\
\label{eq:Z-Sh}
\end{eqnarray}
with the following boundary conditions
\begin{equation}
  \Psi_{x \to -\infty} = \left( \begin{array} {ll}
     1\\
     0
    \end{array} \right) e^{-i\lambda x},\
   \Psi_{x \to \infty} = \left( \begin{array} {ll}
     a(\lambda)\, e^{-i\lambda x} \\
     b(\lambda)\, e^{i\lambda x}
    \end{array} \right).
\label{eq:BC}
\end{equation}
Here $\Psi(x)$ is an eigenvector, $\lambda$ is an eigenvalue,
$a(\lambda)$ and $b(\lambda)$ are the scattering coefficients, and a
star means a complex conjugate. The number $N$ is equal to the
number of poles $\lambda_n \equiv \xi_n + i \eta_n$, where $n = 1, \dots,
N$, and $\eta_n > 0$, of the transmission coefficient
$1/a(\lambda)$. Each $\lambda_n$ is invariant on $z$. If all $\xi_n$
are different then $u(x,z)$ at $z \to \infty$ represents a set of
solitons, each in the form of Eq.~(\ref{eq:Solit}) with $\eta =
\eta_n$ and $\xi = \xi_n$. If real parts of several $\lambda_n$ are
equal then a formation of a neutrally stable bound state of solitons
is possible.

   The solution of the Zakharov-Shabat problem (\ref{eq:Z-Sh}) with
the potential (\ref{eq:IC}) is written as
\begin{eqnarray}
  &&a(\lambda)=
  e^{i (\lambda+\nu_1) w_1}  e^{i (\lambda+\nu_2) w_2}
\nonumber\\
  && \ \ \ \left\{  \left[
    \cos(k_1 w_1) - i {(\lambda +\nu_1) \over k_1} \sin(k_1 w_1)
    \right] \times \right.
\nonumber\\
  && \ \ \ \left[
    \cos(k_2 w_2) - i {(\lambda +\nu_2) \over k_2} \sin(k_2 w_2)
    \right] -
\nonumber\\
  &&\ \left.
  {Q_1^{*} Q_2 \over k_1 k_2}
  e^{-2 i (\lambda+\nu_1) x_2}\,  e^{2 i (\lambda+\nu_2) x_3}
  \sin(k_1 w_1)\, \sin(k_2 w_2)  \right\},
\label{eq:Coefa}
\end{eqnarray}
\begin{eqnarray}
  && b(\lambda) =
  e^{i (\lambda+\nu_1) w_1}  e^{-i (\lambda+\nu_2) w_2}
\nonumber\\
  &&\ \ \ \left\{
    -{Q_1^{*} \over k_1}   e^{-2i(\lambda+\nu_1) x_2} \sin(k_1 w_1)
  \right.
\nonumber\\
  &&\ \ \ \left[
    \cos(k_2 w_2) + i {(\lambda +\nu_2) \over k_2} \sin(k_2 w_2)
  \right] -
\nonumber\\
  &&\ \ \
  {Q_2^{*} \over k_2}  e^{-2i(\lambda+\nu_2) x_3} \sin(k_2 w_2)
\nonumber\\
  &&\ \ \ \left. \left[
    \cos(k_1 w_1) - i {(\lambda +\nu_1) \over k_1} \sin(k_1 w_1)
  \right]  \right\}
\label{eq:Coefb}
\end{eqnarray}
where $k_j = [(\lambda+ \nu_j)^2 + |Q_j|^2]^{1/2}$.

   Since the linear operator in (\ref{eq:Z-Sh}) is not Hermitian,
complex eigenvalues are possible even for real $u(x,0)$ (e. g. see
Section~\ref{Sect:Equal}). Though this is an obvious fact, ``an
interesting ``folklore'' property seems to have arisen in the
literature over the last 25 years, namely, that only pure imaginary
EVs (eigenvalues) can occur for symmetric real valued potentials''
\cite{Klau}. As demonstrated below, the statement [Theorem (III) in
\S 2] of paper \cite{Sats}, that claims this result, is incorrect.
An existence of eigenvalues with non-zero real parts for
Zakharov-Shabat problem with pure real potential was first shown in
paper \cite{Abdul}.

   Equations (\ref{eq:Coefa},\ref{eq:Coefb}) represent a general
solution of the scattering problem (\ref{eq:Z-Sh},\ref{eq:BC}) with
initial condition (\ref{eq:IC}). Applications of these equations to
particular cases of the pulse interaction are considered in the next
section.

\section{Results}
\label{Sect:Res}

\subsection{Interaction of in-phase pulses with equal amplitudes}
\label{Sect:Equal}

\subsubsection{Properties of eigenvalues}

  Here we analyze a simple case of two real pulses, separated by a
distance $L \equiv x_3 -x_2$, with zero detuning, i.e. $Q_1 = Q_2 =
Q_0$, $w_1 = w_2 \equiv w$, and  $\nu_1= \nu_2= 0$, where $Q_0$ is
real. Then using Eq.~(\ref{eq:Coefa}), the equation for discrete
spectrum is written as
\begin{equation}
  F(\lambda, Q_0, w) \pm  {Q_0 \over k}
    e^{i \lambda L}\  \sin(k w)\ = 0 \ ,
\label{eq:EV1}
\end{equation}
where $F(\lambda, Q, w) \equiv \cos(k\, w) - i \lambda \sin(k\,
w)/k$, and $k= (\lambda^2 + Q^2)^{1/2}$. Note that $F(\lambda,Q_0,w)
= 0$ determines the discrete spectrum for a single box with zero
detuning \cite{Manak}. Therefore the second term in
Eq.~(\ref{eq:EV1}) can be associated with the result of nonlinear
interference. Recall also that for a single box with amplitude $Q_0$
and width $w$, the number $N_{SB}$ of emerging solitons is
determined as \cite{Zakh} $N_{SB}= {\rm int} (Q_0 w/\pi+1/2)$, where
${\rm int}()$ means an integer part. Results for the two boxes are
reduced to those for a single box in limiting cases $L=0$ and $L=
\infty$.

   As shown by Klaus and Shaw \cite{Klau}, the Zakharov-Shabat
problem with a ``single-hump'' real initial condition admits pure
imaginary eigenvalues only, i.e. solitons with zero velocity. We
show that the case of two pulses provides much richer dynamics.

\begin{figure}
\centerline{\psfig{figure=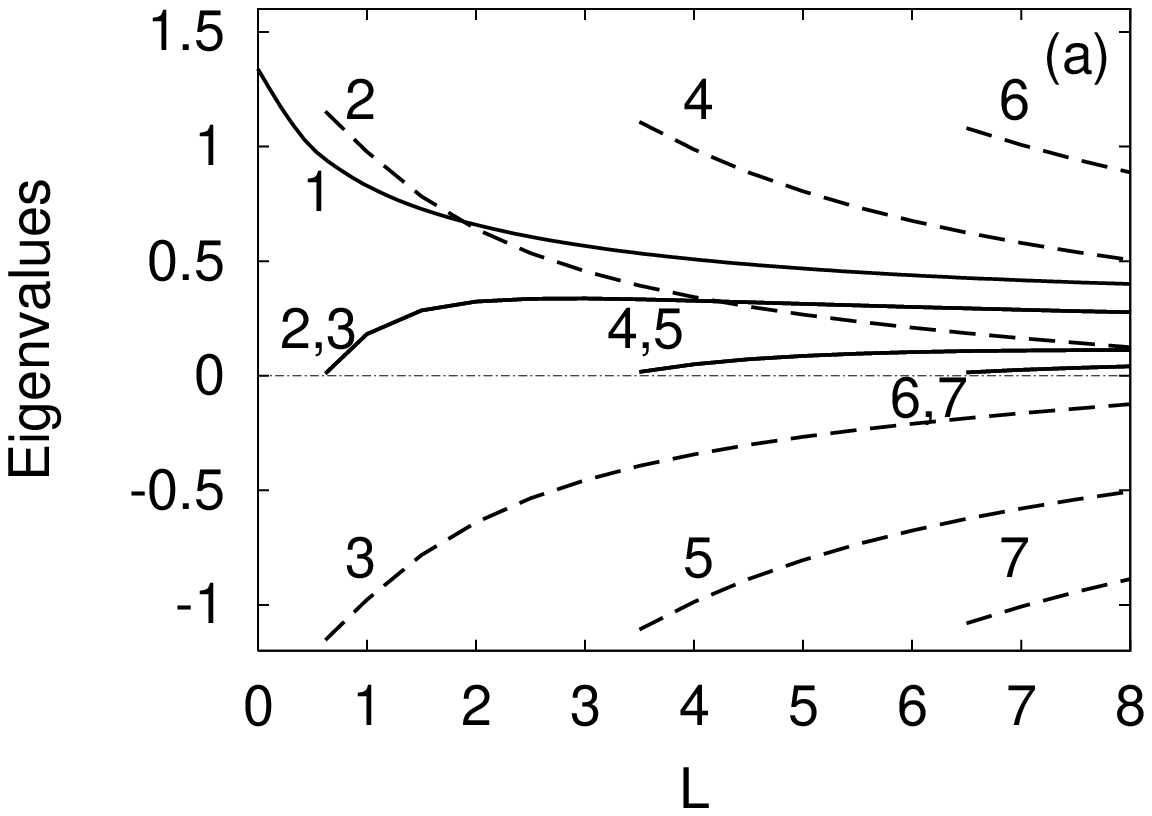,height=5.5cm} \ \ \
\psfig{figure=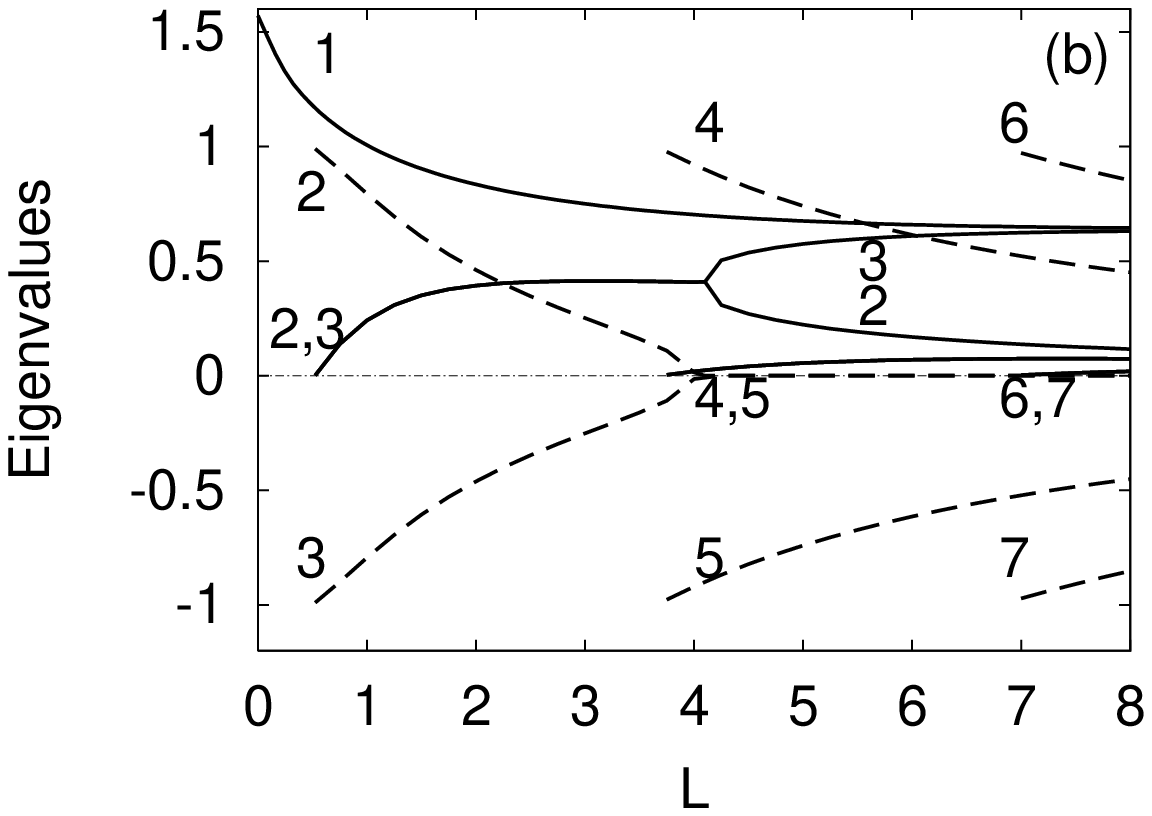,height=5.5cm}}
\centerline{ \psfig{figure=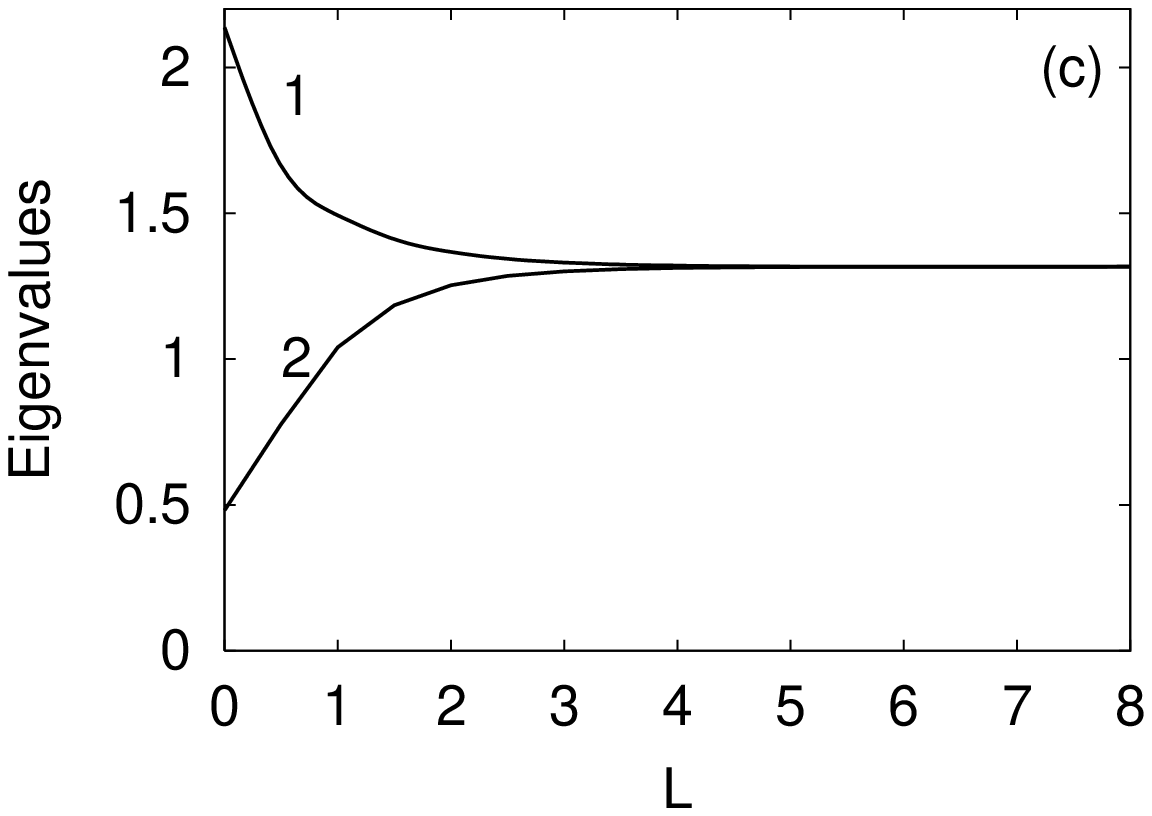,height=5.5cm}}

\caption{
In-phase pulses: The dependence of real (dashed  lines) and
imaginary (solid lines) parts of $\lambda_n$ on the separation
distance, $w= 1$.  Numbers near lines corresponds to $n$. (a)
$Q_0= 1.8$, (b) $Q_0= 2.0$, (c) $Q_0= 2.5$. }

\label{f1-ev1}
\end{figure}

   Let us now compare the properties of eigenvalues at different
$S\equiv Q_0 w $ (Fig.~\ref{f1-ev1}). In Fig.~\ref{f1-ev1}, as well
as in subsequent figures of the paper, all variables are
dimensionless. In the first two cases, $S= 1.8$ and $S=2.0$, there
is one soliton at $L=0$ and there are two solitons at $L= \infty$,
while in the case $S= 2.5$ there are two solitons in both limits.
The dependence of eigenvalues on $L$ at $S=2.5$ is obvious, while
that at $S= 1.8$ and $2.0$ looks unexpected. Firstly, the number of
solitons at intermediate $L$ is larger then that in the limits $L=0$
and $L=\infty$. Secondly, the two real boxes lead to eigenvalues
with non-zero real part. Third, for $S= 2.0$ there is a ``fork''
bifurcation at $L= L_{F}\approx 4.1$, when two eigenvalues coincide.
At larger $L$ three pure imaginary eigenvalues constitute a
three-soliton state, so that the limiting two-soliton case at $L \to
\infty$ is realized as a limit of a three-soliton solution with an
amplitude of the third soliton tending to zero.

\begin{figure}
\centerline{
  \psfig{figure=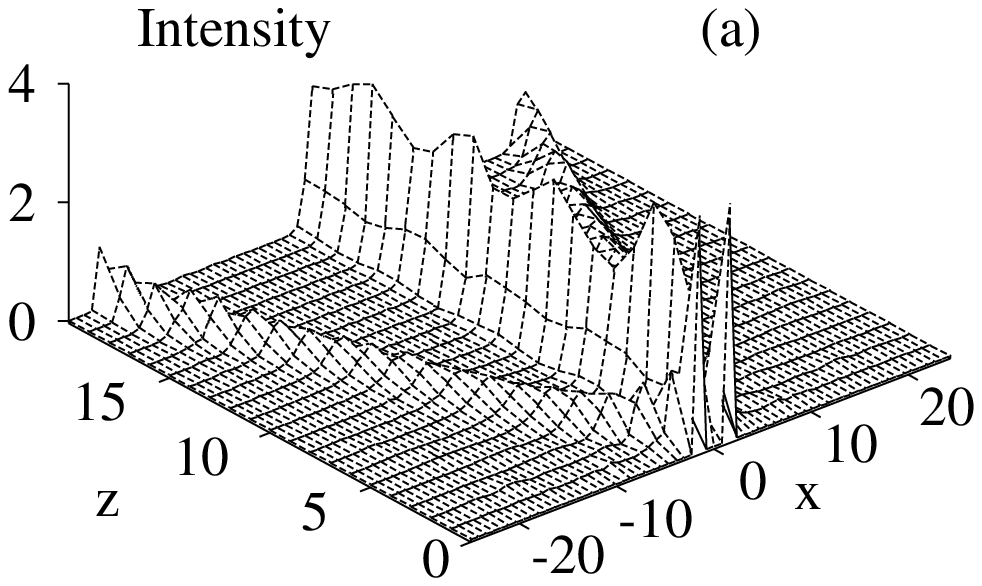,height=6.5cm}
  \psfig{figure=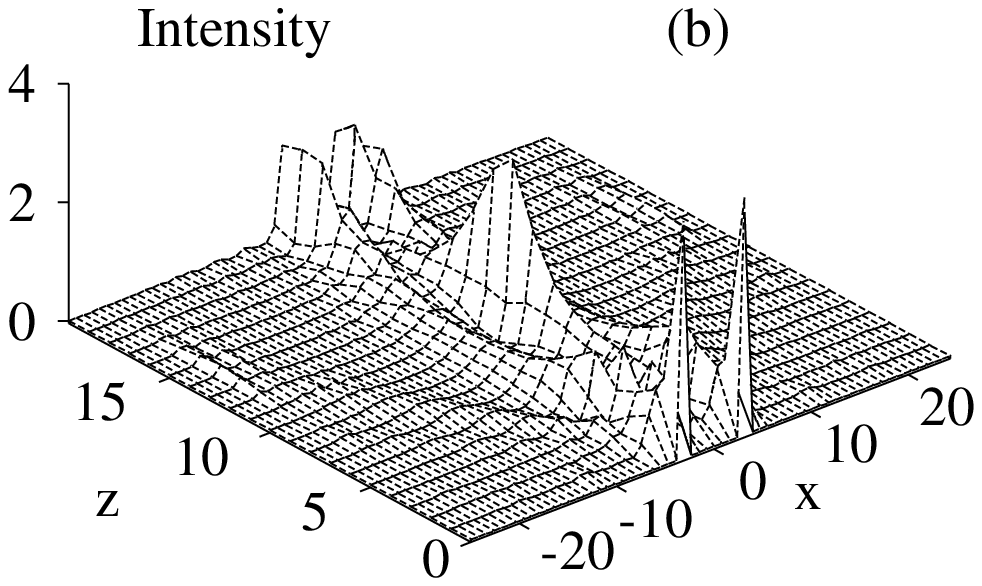,height=6.5cm}
}
\caption{ Evolution of two rectangular pulses, $Q_0= 2,
w= 1$. (a) One fixed soliton and two moving solitons at $L=2$.
(b) Three-soliton state at $L=5$.
}

\label{f2-dyn}
\end{figure}

   Results of numerical simulations of the NLS
equation~(\ref{eq:NLS}) agree with analysis of Eq.~(\ref{eq:EV1}).
For example, as shown in Fig.~\ref{f2-dyn},  in accordance with
Fig.~\ref{f1-ev1}b there are one fixed and two moving solitons at $S
= 2.0$ and $L = 2$, and there are a three-soliton state and two moving
solitons at $S = 2.0$ and $L=5$. Note that an appearance of moving
solitons and multi-soliton states is not related to the rectangular
form of initial pulses. For example, an initial condition $u(x,0) =
0.7 [{\rm sech} (x+2.5) + {\rm sech} (x-2.5)]$ also results in
moving solitons.

   Below we discuss in details the behaviour of the eigenvalues,
namely we find a threshold of appearance of new roots, estimate a
number of emerging solitons, and calculate a threshold for the
``fork'' bifurcation. It should be mentioned that eigenvalues with
non-zero real part do not exist only at $S= [3\pi/4,3.3]$ and $S=
[7\pi/4,5.51]$ (see Section~\ref{Sect:App}), so that the dependence
at $S = 2.5$ is rather an exception than a general rule. This
results allows to understand why moving solitons are not observed in
interaction of near-soliton pulses with area $S \approx \pi$.

\subsubsection{Appearance of new eigenvalues}
\label{Sect:App}

   Solving numerically Eq.~(\ref{eq:EV1}), one can conclude that
new eigenvalues penetrate to the upper half-plane of $\lambda$ in
pairs by crossing the real axis. Therefore, the bifurcation
parameter  can be found from Eq.~(\ref{eq:EV1}), assuming that
$\lambda= \beta$, where $\beta$ is real:
\begin{eqnarray}
  {\rm cotan}\; y &=& \pm \frac{\sqrt{2 S^2 -y^2}}{y} \ ,
\label{eq:Real1} \\
  \beta  &=& \pm Q_0 \sin(\beta L) \ .
\label{eq:Real2}
\end{eqnarray}
Here $y = \kappa w $, $\kappa = (\beta^2 + Q_0^2)^{1/2}$, and the
signs are taken such that $\tan(y) \tan(\beta L) < 0$ is satisfied.
As follows from the definition of $y$ and Eq.~(\ref{eq:Real1}),
one has $S \leq y < 2S$.

    Analysis of Eqs.~(\ref{eq:Real1},\ref{eq:Real2}) results in the
following conclusions:

(i) As follows from Eq.~(\ref{eq:Real1}), the number $N_{PP}$ of the
penetration points depends only on $S$ and is determined from:
\begin{eqnarray}
  N_{PP}&=& 4(m-n+1) - 2 \theta
    \left[ S-  \left( n+{1 \over 4} \right) \pi \right] -
\nonumber \\
&&  2 \theta \left[ S- \left( n+{3 \over 4} \right) \pi \right] -
   4 \theta ( S_{m} - S) \qquad {\rm for}\ \  S \geq 3 \pi/4 \ ,
\label{eq:PP}
\end{eqnarray}
where $m= {\rm int}(\sqrt{2}\ S/\pi)$, $n= {\rm
int}(S/\pi)$, $\theta(x)$ is the Heavyside function, and $S_{m}$
is a root of
\begin{equation}
  \tan (\sqrt {2 S_m^2-1}) = \sqrt {2 S_m^2-1} \ ,
\label{eq:Thr}
\end{equation}
which satisfies $ m\pi \leq (2S_{m}^2-1)^{1/2} < (m+1)\pi$. It is
easy to find that $N_{PP}=0$ for $S < \pi/4$\, and $N_{PP}=2$ for $
\pi/4 < S < 3\pi/4$. Equation (\ref{eq:Thr}) defines such values of
$S= S_m$, when the right hand side of Eq.~(\ref{eq:Real1}) with plus
sign touches ${\rm cotan} y$ curve. All penetration points
$\beta_j$, where $j =1, \dots, N_{PP}$, are symmetrically situated
with respect to $\beta= 0$.

(ii) All roots $|\beta_j| \leq Q_0\ $, which follows from $2S^2-y^2
\geq 0$.

(iii) For every $\beta_j$, Eq.~(\ref{eq:Real2}) defines
the separation distance $L= L_{C}$, when eigenvalues cross the real
axis.

(iv) As follows from Eq.~(\ref{eq:Real2}) there is an
{\em infinite} number of thresholds $L_C$ for a given $\beta_j$.
However, the total number of eigenvalues in the upper half-plane of
$\lambda$ is, most probably, finite, because for some $L_C$
eigenvalues pass to the upper half-plane, and for other $L_C$
eigenvalues go to the lower half-plane. The direction of eigenvalue
motion is defined by the derivative $d\lambda / d L$ at $\lambda=
\beta_j$.

\begin{figure}
\centerline{
  \psfig{figure=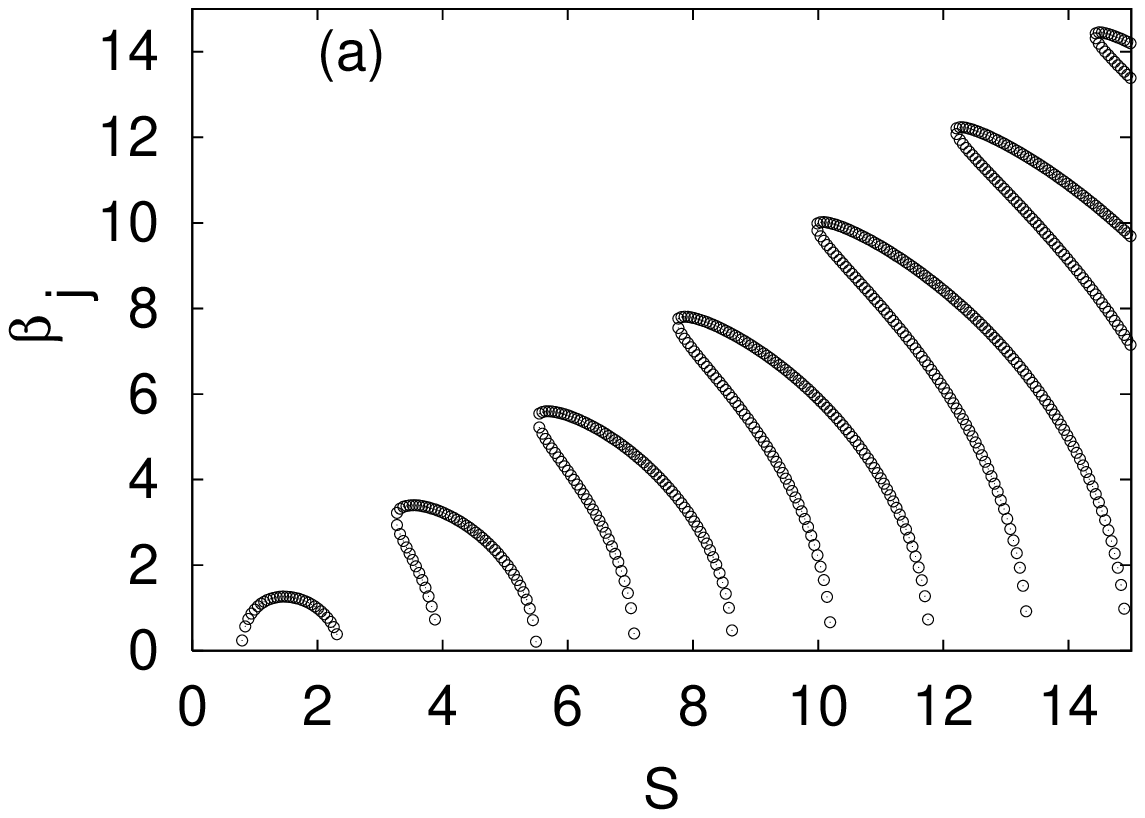,height=5.5cm}
  \psfig{figure=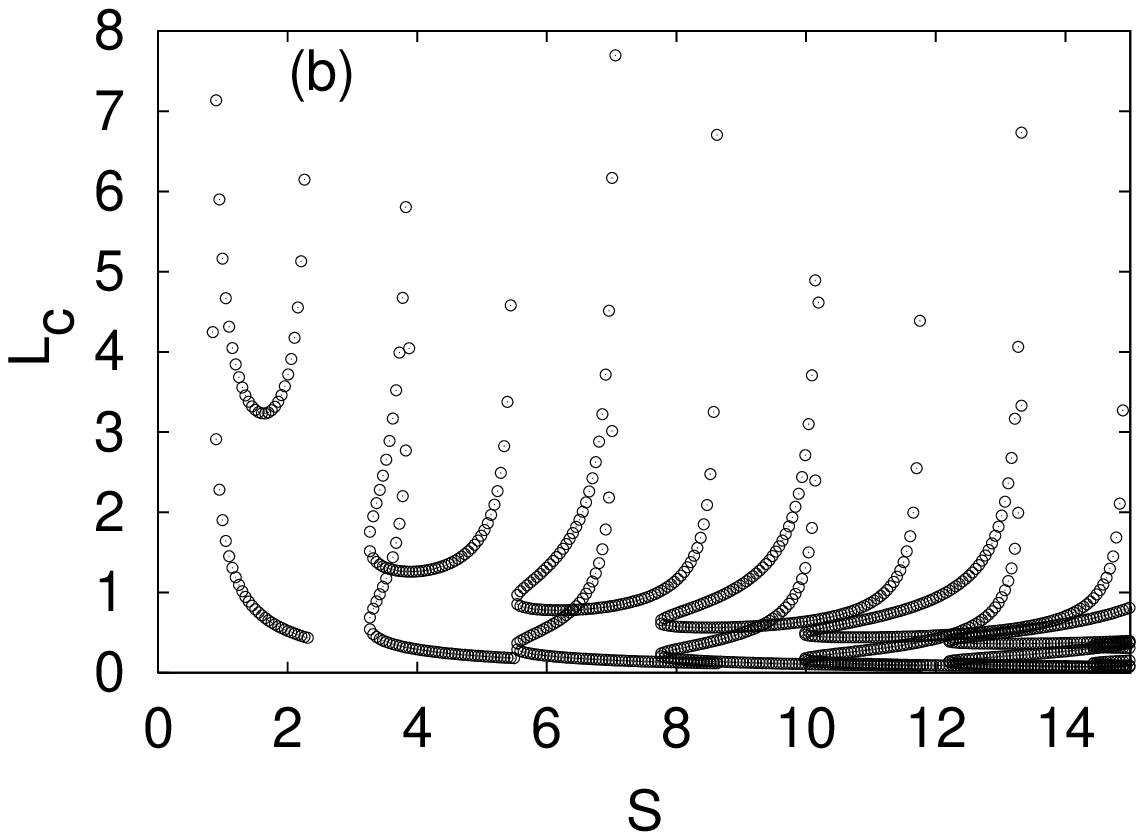,height=5.5cm}
}
\caption{ (a) The dependence of $\beta_j$ on $S$.
(b) Threshold $L_C$, when eigenvalues cross the real
axis of $\lambda$-plane, as a function of $S$.
}

\label{f3-crs}
\end{figure}

   The positions of penetration points, $\beta_j$ as a function of
$S$ is shown in Fig.~\ref{f3-crs}a, where only positive $\beta_j$
are presented. As follows from Eq.~(\ref{eq:Real1}) the number
$N_{PP}$ decreases by two, when $S$ passes $(2l+1)\pi/4$, where $l=
1, 2 \dots$, and $N_{PP}$ increases by four, when $S$ exceeds $S_m$
[see Eq.~(\ref{eq:Thr})]. Therefore one can obtain that
Eq.~(\ref{eq:Real1}) has no roots only at $S=[3\pi/4,S_2]$ and at
$S=[7\pi/4,S_3]$, where $S_2 \approx 3.26$ and $S_3 \approx 5.51$
are found from Eq.~(\ref{eq:Thr}). This property is clearly seen in
Fig.~\ref{f3-crs}. The dependence of $L_C$ on $S$ is presented in
Fig.~\ref{f3-crs}b. Only the thresholds, such that $\beta_j L_C =
[0,2\pi]$, are shown for each $\beta_j$.

\subsubsection{Thresholds of ``fork'' bifurcation}

   Here we analyze a bifurcation, when a pair of complex eigenvalues
becomes pure imaginary, e.g. $L_{F} \approx 4.1$ in
Fig.~\ref{f1-ev1}b. The equation which determine pure imaginary
eigenvalues can be obtained from Eq.~(\ref{eq:Coefa}) with ${\rm
Re}[\lambda] =0$, i.e. $\lambda = i \gamma$:
\begin{equation}
  {\rm cotan}\, y =  \frac{-\sqrt{S^2-y^2}
     \pm S \exp[-\sqrt{S^2-y^2}\; L/w]}{y} \ .
\label{eq:Im}
\end{equation}
Here $y = \kappa \, w\ $, $\kappa = (-\gamma^2 + Q_0^2)^{1/2}$.
It is easy to show that  $\kappa^2$ should be positive (there is no
real solution for $\kappa^2 < 0$). As a
consequence, all pure imaginary eigenvalues satisfy $\gamma_j \leq
Q_0$.

   The value of $L=L_{F}$, when new pure imaginary root of
Eq.~(\ref{eq:Im}) appears, corresponds to the ``fork'' bifurcation.
The bifurcation threshold  $L_F$ can be found from the condition
that the functions, corresponding to the right-hand side of
Eq.~(\ref{eq:Im}), touch  the ${\rm cotan}\, y$ curve.

\begin{figure}
\centerline{
  \psfig{figure=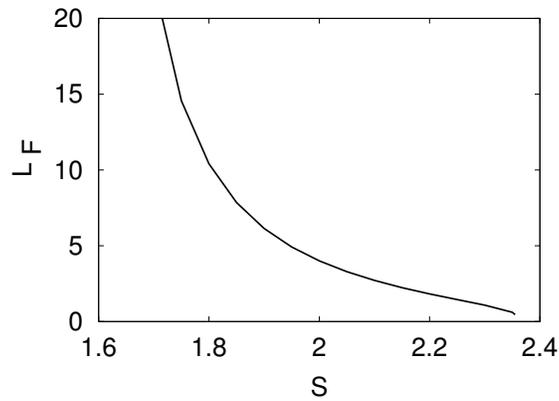,height=5.5cm}
}
\caption{
Threshold $L_F$ of the fork bifurcation as a function of $S$.
}

\label{f4-fork}
\end{figure}

Low bound of the number of the pure imaginary eigenvalues is found
as $N_{LB} = {\rm int}(2S/\pi+1/2)$. This number is an actual number
of the pure imaginary eigenvalues for all $S$, except of the regions
where
\begin{equation}
  \frac{\pi}{2} + \pi l < S < \frac{3\pi}{4} + \pi l,\quad
  l= 0,1,\dots .
\label{eq:Reg}
\end{equation}
If $S$ satisfies Eq.~(\ref{eq:Reg}) then the number of pure
imaginary eigenvalues can be either $N_{LB}$ or $N_{LB}+2$,
depending on whether $L < L_{F}(S)$ or  $L > L_{F}(S)$,
respectively. Therefore the appearance of new pure imaginary
eigenvalues, in other words, the fork bifurcation, is possible only
if $S$ satisfies Eq.~(\ref{eq:Reg}).  Figure~\ref{f4-fork}
represents the dependence of $L_F$ on $S$, where only one interval,
corresponding to $l=1$ in Eq.~(\ref{eq:Reg}), is shown; the
behaviour for $l>1$ is similar. One can calculate that $L_F(S=1.8)
= 10.4$, that is why the fork bifurcation is not seen in
Fig.~\ref{f1-ev1}a.

\subsection{Two out-of-phase pulses with equal amplitudes}

   In this section we study the influence of constant phase shift on
the pulse interaction, i.e. we consider $Q_1 = Q_0 \exp(-i\
\alpha)$, $Q_2 =Q_0 \exp(i\ \alpha)$, where $Q_0$ and $\alpha$ are
real, $w_1 = w_2 \equiv w$, and  $\nu_1= \nu_2= 0$. The non-zero
relative phase shift, $2\alpha$, changes greatly the properties of
the eigenvalues, so that the behaviour presented in
Section~\ref{Sect:Equal} is hard to realize in experiments, because
it is difficult to prepare two pulses exactly in phase. The phase
shift breaks the {\em simultaneous} appearance of a pair of solitons
at $\lambda = \pm \beta_j$, and affects to the ``fork'' bifurcation.

   For $\alpha \neq 0$, the equations for eigenvalues and for
penetration points can be obtained from Eq.~(\ref{eq:EV1}) and
Eqs.~(\ref{eq:Real1},\ref{eq:Real2}) by changing $\lambda L \to
\lambda L + \alpha$ and $\beta L \to \beta L + \alpha$ in the
exponent and sinus functions, respectively.  Therefore the number
and positions of the penetration points are the same as for the case
$\alpha = 0$. As for the threshold $L_C$, it is shifted on the value
$\alpha/\beta_j$, so that $L_C(\alpha) = L_C(\alpha=0) +
\alpha/\beta_j$, where only $L_C \ge 0$ should be taken into
account.

\begin{figure}
\centerline{
  \psfig{figure=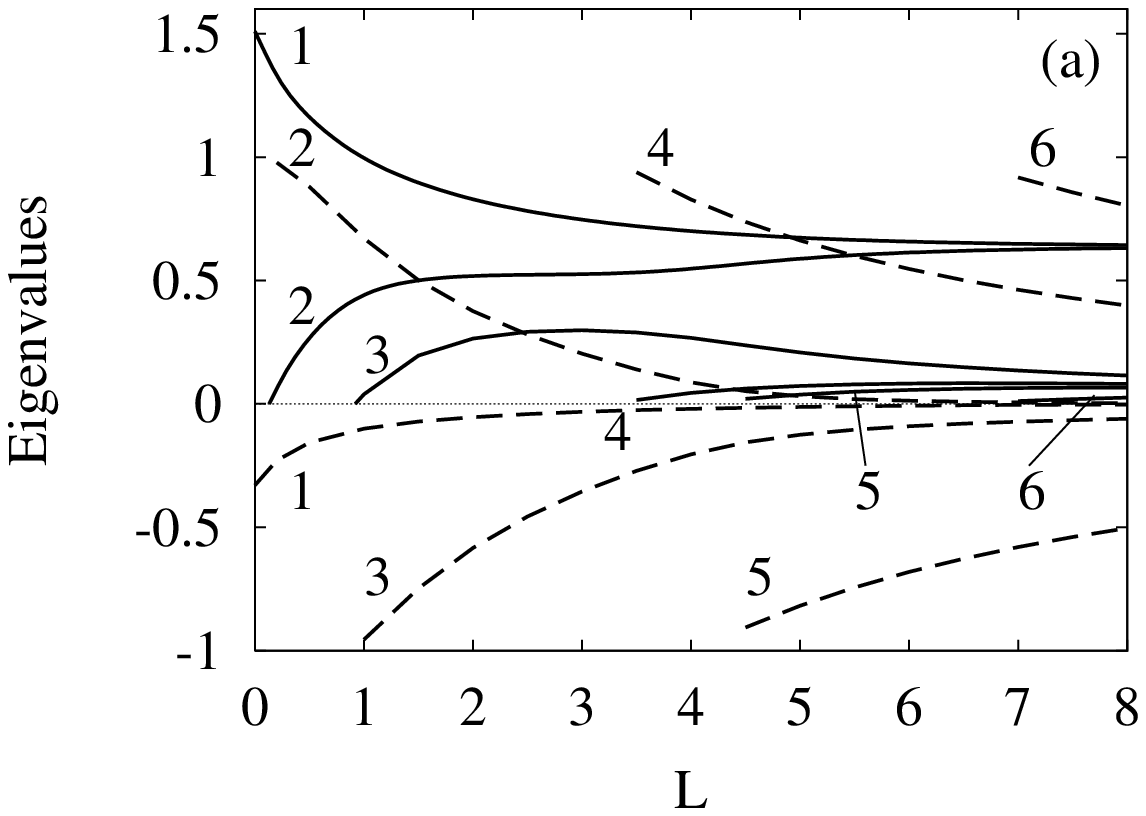,height=5.5cm}
  \psfig{figure=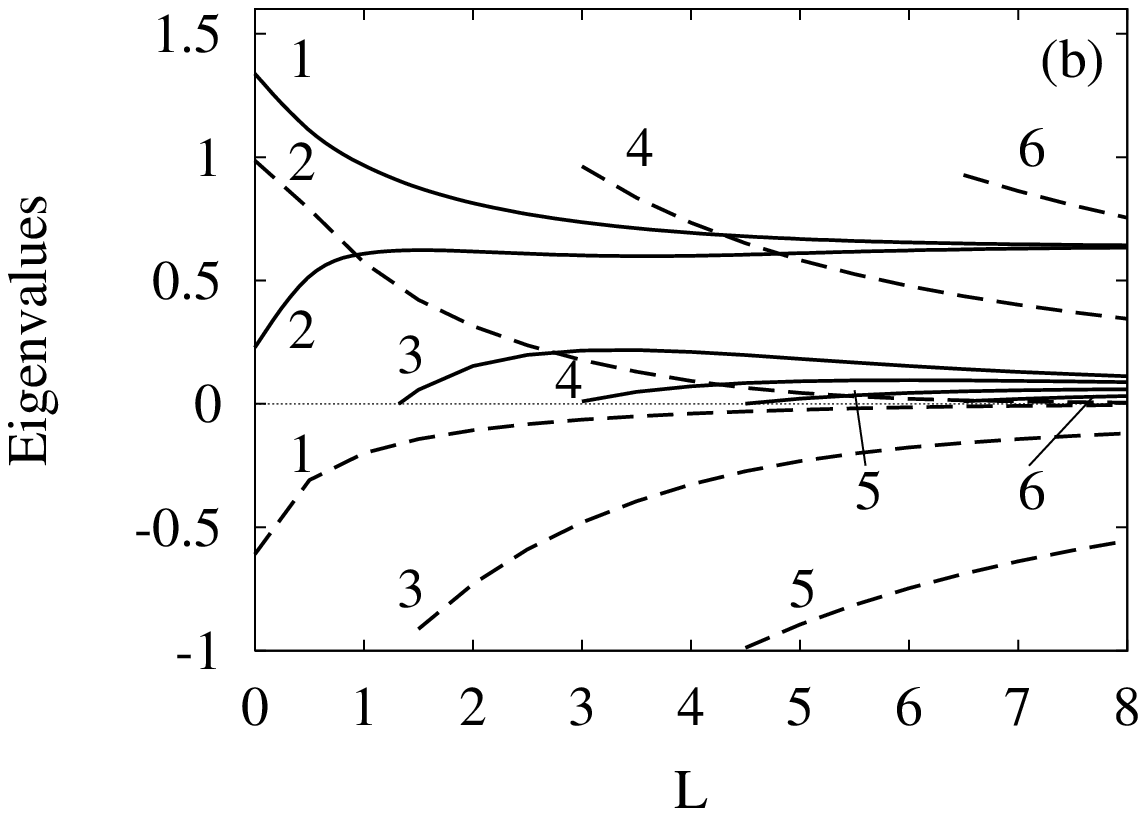,height=5.5cm}
}

\caption{
Out-of-phase pulses: The dependence of real (dashed  lines) and
imaginary (solid lines) parts of $\lambda_n$ on $L$ for  $Q_0= 2, w=
1$. Numbers near lines corresponds to $n$. (a) $\alpha = \pi/8$, (b)
$\alpha = \pi/4$. }

\label{f5-phas}
\end{figure}

   The influence of the phase shift on the distribution of
eigenvalues is shown in Fig.~\ref{f5-phas}. As seen, now new
eigenvalues appear one by one, not in pairs, and, as a concequense,
the ``fork'' bifurcation disappears. Further, the real parts of the
roots do not vanish at finite $L$, but decreases smoothly. This
means that the presence of the phase shift breaks up a multi-soliton
state, which is known to be neutrally stable to perturbations.

   At $L=0$, the phase shift corresponds to the phase jump of a
single pulse. Such a phase jump can result in an appearance of
additional solitons as shown in Fig.~\ref{f5-phas}b. The threshold
of the phase shift $\alpha_{th}$, when the first new soliton appear,
can be found from the condition $\alpha_{\rm th} = |\beta_1|
L_C(\alpha=0)$, where $\beta_1$ is the position of the penetration
point nearest to zero.

\subsection{Two pulses with frequency separation}

   In this section we analyze the initial condition~(\ref{eq:IC})
with the following parameters $Q_1=Q_2=Q_0$, $w_1= w_2 =w$, $-\nu_1 =
\nu_2=\nu$, where $Q_0$ is a real constant. This case models the
wavelength division multiplexing in optical fibers, the case when an
input signal consists of two or more pulses with different
frequencies. Actually, since $Q_0$ can be taken sufficiently large we
consider the interaction of multi-soliton states. The detailed
analysis of the interaction of sech-pulses at different frequencies
is presented in papers \cite{WDM} and in review \cite{Pano}. In
particular, the authors of papers \cite{WDM,Pano} consider the
evolution of a superposition of $N$ solitons with the same position
of the centers, but with different frequencies. As shown in these
works there is a critical frequency separation, above which
$N$ solitons with almost equal amplitudes emerge. Below this
critical value the number of emerging solitons can be not equal $N$
and their amplitudes can appreciable differ from each
other. It was also demonstrated that an introduction of a time shift
between pulses results in decrease of the frequency separation
threshold. The geometry described by Eq.~(\ref{eq:IC}) corresponds
to  the combination of WDM and time-division multiplexing schemes,
therefore our study can give some insight to such a behaviour of the
threshold. Moreover, the authors of works \cite{WDM,Pano} mostly
used the perturbation technique and numerical simulations, while in
the present paper we deal with an exact solution of the
Zakharov-Shabat problem.

\begin{figure}
\centerline{
  \psfig{figure=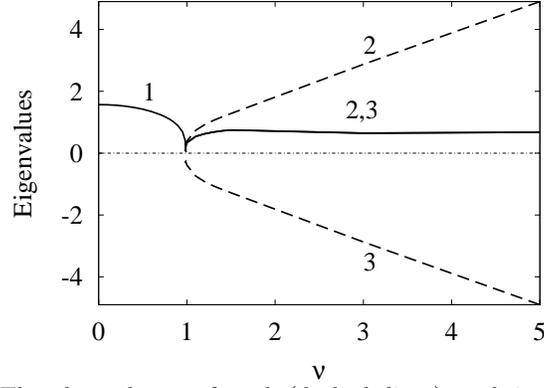,height=5.5cm}
}
\caption{
Single chirped pulse; The dependence of real (dashed  lines) and
imaginary (solid lines) of $\lambda_n$ on $\nu$ for  $Q_0= 2,
w= 1, L = 0$. Numbers near lines corresponds to $n$.
}

\label{f6-chrp}
\end{figure}

   First let us consider the case $L=0$ that correspond to the case
of a single {\em chirped} pulse of width $2w$. The dependence of the
eigenvalues on $\nu$, which plays here the role of a chirp
parameter, is shown in Fig.~\ref{f6-chrp}. At small $\nu$ the
interaction of the pulse components is strong, so that there is one
pure imaginary eigenvalue, or a single soliton with zero velocity.
At larger $\nu$ the frequency difference of the pulse components
results in a repulsion of the components, or a pulse splitting. At
sufficiently large $\nu$ the velocities of emerging solitons tends,
as expected, to $\pm 2\nu$.

   There is also a narrow region of $\nu$, e.g. $\nu=[0.98,0.99]$ on
Fig.~\ref{f6-chrp}, where three solitons, one fixed and two moving
solitons, exist. This region separate two different types of the
evolution of a chirped pulse. The left boundary, which corresponds
to the appearance of new eigenvalues, of the region is found from
the condition similar to that considered in Section~\ref{Sect:App}.
The right boundary is found from $a(\lambda = 0) =0$, which defines
the values of $\nu$ as a function of the other parameters, when
the pure imaginary root disappears.

\begin{figure}
\centerline{
  \psfig{figure=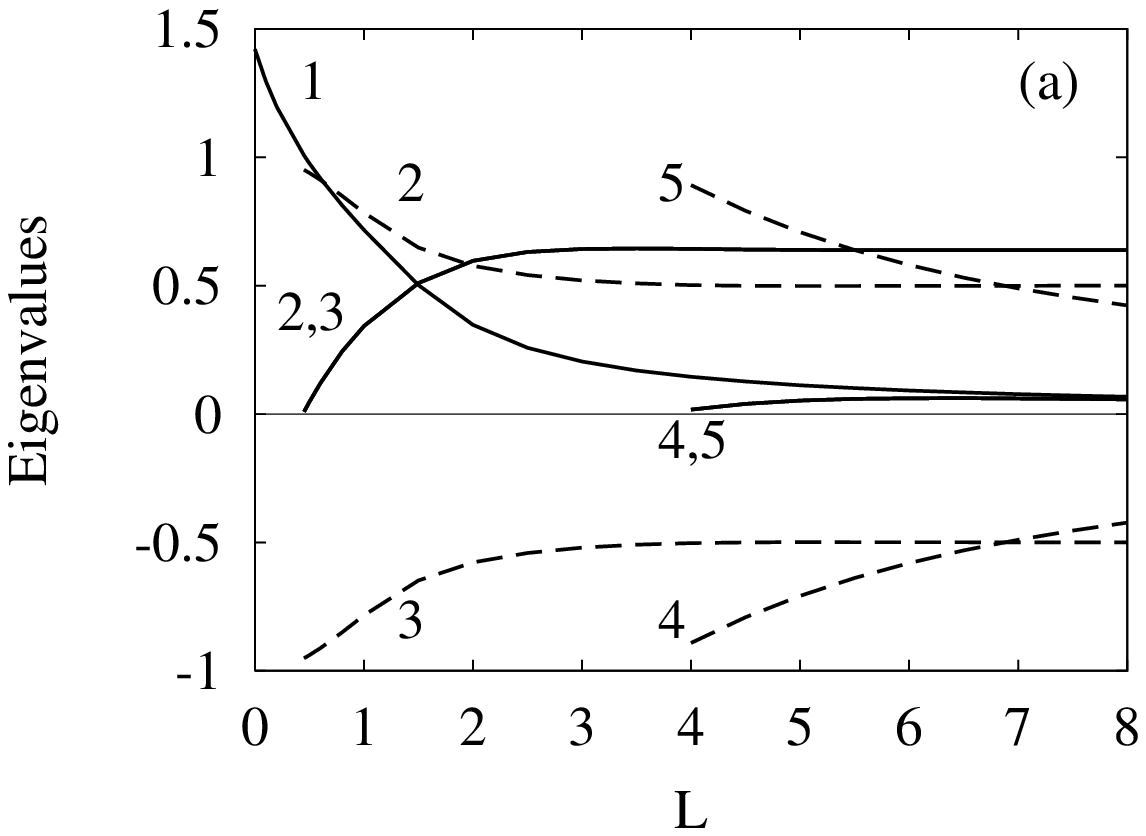,height=5.5cm}
  \psfig{figure=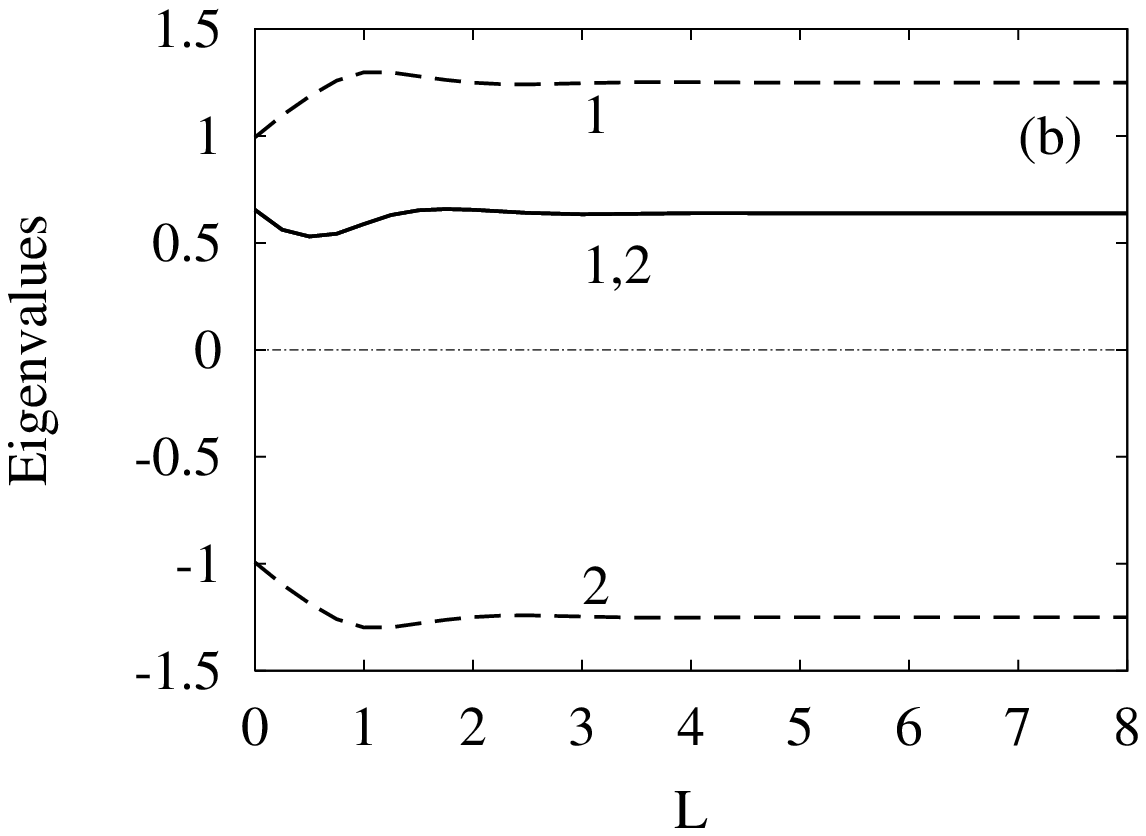,height=5.5cm}
}
\caption{
Pulses with frequency separation: The dependence of real (dashed
lines) and imaginary (solid lines) of $\lambda_n$ on $L$ for $Q_0=
2, w= 1$. Numbers near lines corresponds to $n$. (a) $\nu = 0.5$;
(b) $\nu = 1.25$.
}

\label{f7-fss}
\end{figure}

    The dependence of the spectrum on $L$ is presented in
Fig.~\ref{f7-fss}. At small $\nu$ (Fig.~\ref{f7-fss}a) we see again
an appearance of additional solitons similar to the case $\nu=0$
(Fig.~\ref{f1-ev1}). At larger $\nu$ (Fig.~\ref{f7-fss}b) the
repulsion is so strong that it suppresses the appearance of
small-amplitude solitons. Therefore there is a threshold of the
frequency separation above which the interaction of two pulses is
negligible. This result is in agreement with the conclusions of the
paper \cite{Pano}.

\section{ Conclusion }
\label{Sect:Conc}

  The interaction of two pulses in the NLS model is studied by means
of the solution of the associated scattering problem. The strong
dependence of the dynamics on the parameters of the initial pulses
is shown. For intermediate separation distances $L$ the existence of
additional moving solitons is possible even in the case of two
in-phase pulses with the same frequencies. These additional solitons
can be considered as a result of nonlinear interference of pulses.
The phase shift of two pulses removes a degeneracy
in the behaviour, namely it affects to the symmetry of the
parameters of emerging solitons and results in a break-up of
multi-soliton states peculiar to the in-phase case. It is also
shown that the strong frequency separation suppress the appearance
of additional solitons. The results obtained in the present paper
can be useful for analysis of the transmission capacity of
communication systems and for interpretation of experiments
on the interaction of two laser beams in nonlinear media.
Recently, the generation of up to ten solitons has been
observed experimentally in quasi-1D Bose-Einstein condensate
of $^{7}{\rm Li}$ with attractive  interaction \cite{Streck}.
Our study can be also helpful for interpretation of this experiment.

\acknowledgments

    This researh was partially supported by the Foundation for
Support of Fundamental Studies, Uzbekistan (grant N 15-02) and by
FAPESP (Brasil).

\end{document}